\documentclass[reprint,twocolumn,superscriptaddress,showpacs,floatfix]{revtex4-1}

\usepackage{epsfig}
\usepackage{amsmath}
\usepackage{exscale}
\usepackage{array}
\usepackage{units}
\usepackage{amssymb}
\usepackage{setspace}
\usepackage{bbm}
\usepackage{color}
\usepackage{color}
\definecolor{lightgrey}{rgb}{0.87,0.87,0.87}

\begin{document}

\title{Fractional charge and spin errors in self-consistent Green's function theory}


\author{Jordan J. Phillips\footnote{Corresponding Author:~email@umich.edu}}
\affiliation{Department of Chemistry, University of Michigan, Ann Arbor, Michigan 48109, USA}
\author{Alexei A. Kananenka}
\affiliation{Department of Chemistry, University of Michigan, Ann Arbor, Michigan 48109, USA}
\author{Dominika Zgid}
\affiliation{Department of Chemistry, University of Michigan, Ann Arbor, Michigan 48109, USA}

\begin{abstract}

We examine fractional charge and spin errors in self-consistent Green's function theory within a second-order approximation (GF2). For GF2 it is known that the summation of diagrams resulting from the self-consistent solution of the Dyson equation removes the divergences pathological to second-order M\o ller-Plesset  theory (MP2) for strong correlations. In the language often used in density functional theory contexts, this means GF2 has a greatly reduced {\em fractional spin error} relative to MP2. The natural question then is what effect, if any, does the Dyson summation have on the {\em fractional charge error} in GF2? To this end we generalize our previous implementation of GF2 to open-shell systems and analyze its fractional spin and charge  errors. We find  that like MP2, GF2 possesses only a very small  fractional charge error, and consequently little many electron self-interaction error. This shows that GF2 improves on the critical failings of MP2, but without altering the positive features that make it desirable. Furthermore, we find that GF2 has \emph{both} less fractional charge and fractional spin errors than typical hybrid density functionals as well as random phase approximation with exchange. 

\end{abstract}
\maketitle

\newpage

\section{Introduction}

Self-consistent single-particle electronic structure methods are of great interest because they combine conceptual and computational simplicity while being free of a starting reference bias. Probably the most well known example of this is density functional theory (DFT)\cite{PhysRev.136.B864,*PhysRev.140.A1133}. Often these methods are designed  to satisfy known constraints that an exact electronic structure theory should obey. For example, the total electronic energy should vary linearly in the fractional electron occupancy between integer electron numbers\cite{Perdew1982,fracsie3,OneTwoManySIE2007,XC_OpenFluctuatingNumber_Perdew_pra_2007}, while for a simple one-electron system like the hydrogen atom  the energy should be degenerate with respect to variations in the fractional spin\cite{Cohen2008b}. To the extent that a method disobeys these two exact constraints such a method will display many electron self-interaction  and static correlation errors, respectively.
 To understand how these errors are connected to fractional electron behavior, let us consider the H$_{2}$ molecule in the infinite dissociation limit in two different scenarios:
 
 First let us focus on a {\em fractional charge error}.
 If an extra electron were placed on this system with fractional occupancies on each H atom given by $n_a$, $n_b$, $n_a +n_b =1$, then the net energy change of the system would be $E_{\Delta}=-(n_a A + n_b A)=-A$, where $A$ is the electron affinity of the hydrogen atom.  The energy of each subsystem would vary linearly in the occupation, and the total energy of the entire [H$_{2}$]$^-$ system would be the same regardless if the extra electron were delocalized across both atoms, or localized only on one. If the energy of the subsystem varied nonlinearly, then either the delocalized or localized solution would become unphysically lower in energy depending on whether the curve was convex or concave. This unphysical behavior would be a simple manifestation of many electron self-interaction error.
 
 Let us examine now  a {\em fractional spin error}.  For  neutral singlet H$_{2}$ in the infinite dissociation limit each H atom should have half a spin up and down electron, and the  energy of the singlet should be identical to that of broken-symmetry solutions where  spin up and spin down electrons have localized on different atoms. Therefore a method's failure to yield equivalent energies for one-electron H with fractional or integer spin is equivalent to the failure to describe multireference static correlation energy.

From this simple example it is clear that fractional charge and fractional spin errors are deeply connected to many electron self-interaction and static correlation errors.
 For this reason these errors have been studied extensively, and are known to have severe negative consequences for a given method's  description of  properties that  depend on electron delocalization and static correlation effects\cite{TestsofFunct2007,Cohen2008b,FractionalBandGap_Cohen_prb_2008,*FractionalBandGap_Cohen_prl_2008,DelocErrorErin2008,FracChargeFracSpin_jcp_2011_Johnson}.
 This language of fractional charge and spin  has traditionally been used exclusively within the DFT community to analyze approximate density functionals. However these concepts  have begun to make inroads into other areas such as wavefunction\cite{WavefunctionFractional_Weitao_jcp_2013} and many-body theory\cite{Weitao_MP2_Frac_jctc2009,RPA_failure_Weitao_pra_2012,PPRPA_Weitao_pra_2013,GreenFractionalWeitao_jcp_2013,StaticCorrelationRPAGW_2014_arxiv}, and density matrix theory as well\cite{DensMatTheory_fractional_Ayers_pccp_2009}.
 For example it has  been shown that MP2 (second order M$\o$ller-Plesset\cite{MollerPlesset1934}) possesses relatively little fractional charge error, but displays a massive diverging fractional spin error (which is an alternative way of stating that MP2 diverges for strong correlations)\cite{Weitao_MP2_Frac_jctc2009}. Double-hybrid density functionals that include some MP2 correlation are expected to have a similar fractional electron behavior\cite{DoubleHybrids_ijqc_2014,FracDoubleHybrid_Weitao_jpca_2014}. The Random Phase Approximation (RPA)\cite{RPA1,*RPA2,*RPA3,*RPA_GellMann_Brueckner_1957} on the other hand has minimal fractional spin error, but a severe fractional charge error\cite{RPA_failure_Weitao_pra_2012}. The closely related GW approximation\cite{HedinGWpra1965}  displays a fractional charge error similar to RPA, but a larger fractional spin error\cite{StaticCorrelationRPAGW_2014_arxiv}.
 
\begin{figure}
\includegraphics[width=8.5cm]{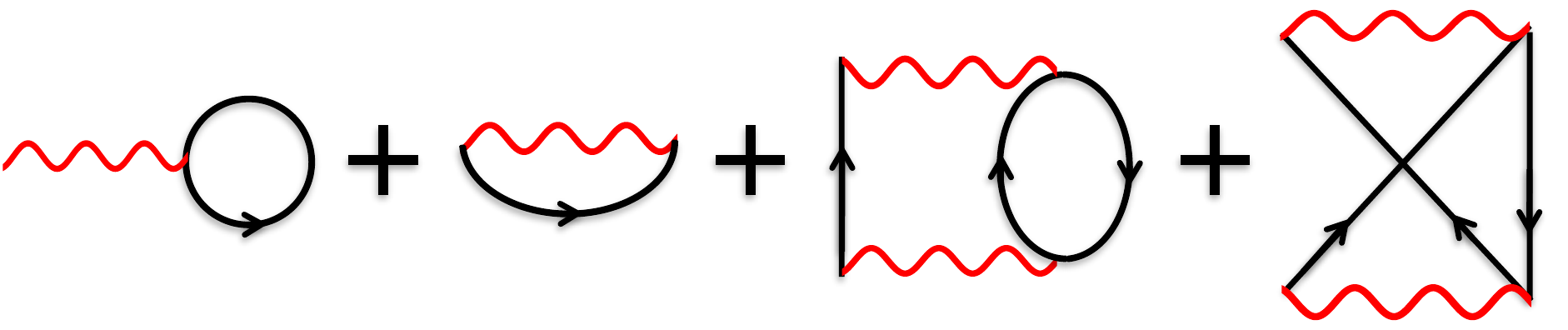}
\caption{Feynman diagrams for the second order self-energy in GF2. Here a red wavy line represents a two electron integral, while a black arrow line represents a Green's function. From left to right the diagrams shown are the first order Hartree and exchange diagrams, and the second order pair bubble and second order exchange.}
\label{fig:diagrams}
\end{figure}

 Motivated by these works, we find it interesting to extend this analysis to self-consistent Green's function theory in a second order approximation  (GF2)\cite{gf2_paper_2014,Dahlenjcp2005}. Similar to MP2, GF2 includes all diagrams to second order in the bare electron-electron interaction, as shown in Figure~\ref{fig:diagrams}, and is therefore exact for one electron systems. However in contrast to MP2 these diagrams are evaluated with self-consistent Green's functions, obtained by iterative solution of the Dyson equation

\begin{equation}
\begin{split}
G(\omega)=G_{0}(\omega)+G_{0}(\omega)\Sigma(\omega)G_{0}(\omega)\\+G_{0}(\omega)\Sigma(\omega)G_{0}(\omega)\Sigma(\omega)G_{0}(\omega)+\cdots \\
=G_{0}(\omega)\biggr(\sum_{n}\bigr(\Sigma(\omega)G_{0}(\omega)\bigr)^{n}\biggr)\\ 
=\bigr[G_{0}(\omega)^{-1}-\Sigma(\omega)\bigr]^{-1}\\
\end{split}
\label{eq:DysonSum}
\end{equation}

\noindent Here $G_{0}(\omega)$ is the Green's function of a non-interacting system, while $\Sigma (\omega)$ is the proper self-energy, which in GF2 is truncated at second order and written as an approximate functional of the Green's function, $\Sigma[G(\omega)]$.  Because of the structure of the Dyson equation, the self-consistent  $G(\omega)$ will contain an infinite order summation of the second order proper self-energy parts, $\Sigma(\omega)$. As we recently showed, this summation of diagrams allows GF2 to give reasonably fine results for strongly correlated systems such as stretched hydrogen lattices\cite{gf2_paper_2014} when MP2 would diverge. In the language of fractional electron errors, this suggests that GF2 improves tremendously over MP2 for fractional spins as a result of the self-consistent infinite order summation. An interesting question that arises then is what effect does this Dyson summation have on the more general fractional electron behavior?  Relative to other methods such as RPA, GW, approximate DFT, and Hartree-Fock (HF), MP2 has only a very small fractional charge error\cite{Weitao_MP2_Frac_jctc2009}, and consequently little many electron self-interaction error (SIE). Ideally  one would hope that GF2 improves on the disastrous fractional spin error of MP2 without deteriorating MP2's impressively small fractional charge error. To investigate this question, here we will generalize our previous GF2 implementation\cite{gf2_paper_2014} to open-shell systems and then investigate its fractional charge and spin behavior.

Before closing this section it should be emphasized that what is  challenging about the fractional charge and fractional spin errors is that any attempt to reduce one error tends to exacerbate the other\cite{SIE_SPE_localhybrid_scuseria_jcp_2010,SIE_and_StaticCorrelation_Henderson_2010,RPA_failure_Weitao_pra_2012}. For example, a  semilocal DFT functional (such as BLYP\cite{B88,LYP}, or PBE\cite{PBE}) will tend to have a  large fractional charge error but a relatively smaller fractional spin error. On the other end of the extreme Hartree-Fock will have significantly less fractional charge error but a much greater fractional spin error. Any hybrid of these two (B3LYP\cite{B3PW91,LYP} or PBEh\cite{PBE,PBE0}, for example) will essentially trade one error for the other to the extent that the HF-type exchange is admixed in place of DFT exchange. What is worth noting is that, in the language of hybrid DFT, the Fock matrix in GF2 contains full HF-type exchange (which in Green's function theory is usually referred to as  first order exchange) yet we will show GF2 yields \emph{both} less fractional charge and fractional spin error than HF, B3LYP, and PBEh. This unique result comes about from a combination of the Dyson summation with including all diagrams to second order.

\section{spin unrestricted GF2 theory}

To study open-shell systems we generalize $\mathbf{G}(\omega)$ to have two spin blocks

\begin{equation}
\mathbf{G}=
\begin{bmatrix}
\mathbf{G}_{\alpha} & \mathbf{0} \\
\mathbf{0} & \mathbf{G}_{\beta}
\end{bmatrix}
\end{equation}

\noindent where the spin-up and spin-down blocks are given by

\begin{equation}
\begin{split}
\mathbf{G}_{\sigma}(\omega)=\bigr[(\mu_\sigma+\omega)\mathbf{S}-\mathbf{F}_{\sigma}-\mathbf{\Sigma}_\sigma(\omega)\bigr]^{-1} ~,~~\sigma =\alpha ,\beta
\end{split}
\label{eq:G}
\end{equation}

\noindent The off-diagonal spin-blocks of $\mathbf{G}(\omega)$ here are identically $\mathbf{0}$, meaning we do not allow for the possibility of spin-flips, and our solutions are constrained to be eigenstates of $\hat{S}_{z}$. In Eq.~\ref{eq:G} $\mathbf{S}$ and $\mathbf{F}_\sigma$ are the overlap and Fock matrices, $\mathbf{\Sigma}_{\sigma}(\omega)$ is the self-energy, $\mu_\sigma$ is the chemical potential, and $\omega$ is an imaginary frequency. By introducing  $\mu_\alpha$, $\mu_\beta$ as separate chemical potentials we can allow for different numbers of electrons in the respective correlated density matrices, $\mathbf{P}_\alpha$, $\mathbf{P}_\beta$, which are given by $\mathbf{P}_{\sigma}=-\mathbf{G}_{\sigma}({\scriptstyle\tau=1/k_BT})$, $\sigma=\alpha,\beta$, where $\mathbf{G}_\sigma(\tau)$ is the Green's function fast Fourier transformed (FFT) to the imaginary time domain, and $1/k_BT$ is the inverse-temperature. The expression for $\mathbf{F}_{\sigma}$ is the standard result from spin-unrestricted HF theory, 

\begin{equation}
\begin{split}
F^{\alpha}_{ij}=h_{ij}+\underset{kl}{\sum}(P^{\alpha}_{kl}+P^{\beta}_{kl})\textnormal{v}_{ijkl}-P^{\alpha}_{kl}\textnormal{v}_{iklj} ~,\\
F^{\beta}_{ij}=h_{ij}+\underset{kl}{\sum}(P^{\alpha}_{kl}+P^{\beta}_{kl})\textnormal{v}_{ijkl}-P^{\beta}_{kl}\textnormal{v}_{iklj} ~.
\end{split}
\label{eq:Fock}
\end{equation}

\noindent However, unlike HF theory the density-matrices that enter this expression are those obtained from the Green's function and thus include electron correlation effects from solving the Dyson equation. This covers the electron-electron interaction from zeroth through first order (the first order diagrams in Figure~\ref{fig:diagrams} are described by the HF mean-field). At second order in GF2 the electron-electron interaction is described by the frequency dependent self-energy, which is given in the imaginary time domain as


\begin{widetext}
\begin{equation}
\begin{split}
\Sigma^{\alpha}_{ij}(\tau)=\underset{klmnpq}{\sum}-G^{\alpha}_{mn}(\tau)G^{\alpha}_{kl}(\tau)G^{\alpha}_{pq}(-\tau)\textnormal{v}_{imqk}\bigr(\textnormal{v}_{lpnj}-\textnormal{v}_{nplj}\bigr) \\
-G^{\alpha}_{mn}(\tau)G^{\beta}_{kl}(\tau)G^{\beta}_{pq}(-\tau)\textnormal{v}_{imqk}\textnormal{v}_{lpnj} ~,\\
~\\
\Sigma^{\beta}_{ij}(\tau)=\underset{klmnpq}{\sum}-G^{\beta}_{mn}(\tau)G^{\beta}_{kl}(\tau)G^{\beta}_{pq}(-\tau)\textnormal{v}_{imqk}\bigr(\textnormal{v}_{lpnj}-\textnormal{v}_{nplj}\bigr) \\
-G^{\beta}_{mn}(\tau)G^{\alpha}_{kl}(\tau)G^{\alpha}_{pq}(-\tau)\textnormal{v}_{imqk}\textnormal{v}_{lpnj} ~,
\end{split}
\label{eq:Sigma}
\end{equation}
\end{widetext}

\noindent  The reasoning for this expression is that at second order a spin-up (down) electron can have a pair-bubble interaction with both spin-up and down electrons, yet the second-order exchange can only proceed between like-spin electrons because we do not allow for the possibility of spin-flips. Once $\mathbf{\Sigma}_{\sigma}(\tau)$ has been built it can be FFT to the frequency domain, and then we can rebuild $\mathbf{G}_{\sigma}(\omega)$ with Eq.~\ref{eq:G}. Looking  at Equations \ref{eq:G}, \ref{eq:Fock}, and \ref{eq:Sigma}, the spin-up and spin-down Green's functions are coupled by the fact that $\mathbf{G}_{\beta}$ appears in the expression for $\mathbf{\Sigma}_{\alpha}$ (likewise with $\mathbf{G}_{\alpha}$ for $\mathbf{\Sigma}_{\beta}$), and that $\mathbf{G}_{\beta}$ contributes to $\mathbf{F}_{\alpha}$ through $\mathbf{P}_{\beta}=-\mathbf{G}_{\beta}({\scriptstyle\tau=1/k_BT})$ (likewise with $\mathbf{G}_{\alpha}$, $\mathbf{P}_{\alpha}$ and  $\mathbf{F}_{\beta}$). Therefore Equations \ref{eq:G}, \ref{eq:Fock} and \ref{eq:Sigma} will need to be solved self-consistently, and at every iteration $\mu_{\alpha}$ and $\mu_{\beta}$ will need to be adjusted to give the desired number of $\alpha$ and $\beta$ electrons. To  start this self-consistent procedure we use a HF Green's function generated by output from the Dalton program\cite{Dalton}.
 Note  when $\mathbf{G}_{\alpha}=\mathbf{G}_\beta$, then $\mathbf{\Sigma}_{\alpha}=\mathbf{\Sigma}_{\beta}$ and Eq.~\ref{eq:Sigma} reduces to the familiar expression for spin-restricted GF2\cite{Dahlenjcp2005,gf2_paper_2014}.

The energy is evaluated as

\begin{equation}
\begin{split}
E=\frac{1}{2}\textnormal{Tr}\bigr[(\mathbf{h}+\mathbf{F}_{\alpha})\mathbf{P}_\alpha+(\mathbf{h}+\mathbf{F}_{\beta})\mathbf{P}_\beta\bigr] \\
+k_BT\underset{n}{\sum}\textnormal{Re}\Bigr[\textnormal{Tr}\bigr[\mathbf{G}_{\alpha}(\omega_{n})\mathbf{\Sigma}_{\alpha}(\omega_{n})+\mathbf{G}_{\beta}(\omega_{n})\mathbf{\Sigma}_{\beta}(\omega_{n})\bigr] \Bigr]
\end{split}
\end{equation}

\noindent where $\omega_n$ is a Matsubara frequency, $\omega_{n}=(2n+1)\pi k_BT$. This can be understood as essentially a  spin-unrestricted HF-like energy expression supplemented with a frequency dependent correlation contribution from GF2. However it should be understood that all quantities are evaluated using the correlated $\mathbf{P}_{\sigma}$ as obtained from $\mathbf{G}_{\sigma}$.

Because the GF2 approximation includes in the proper self-energy all exchange and Coulomb type diagrams to second order, it is by construction exact for one-electron systems, \emph{i.e.} it is one-electron self-interaction and self-correlation free. Less clear however  is its  many electron self-interaction error (SIE) for general systems. To this end in the following we investigate the fractional electron behavior of GF2 for several archetypical cases, and compare to standard density functional theory calculations ran with Gaussian 09\cite{g09}.

\section{Results}

\begin{figure}
\includegraphics[width=8.5cm]{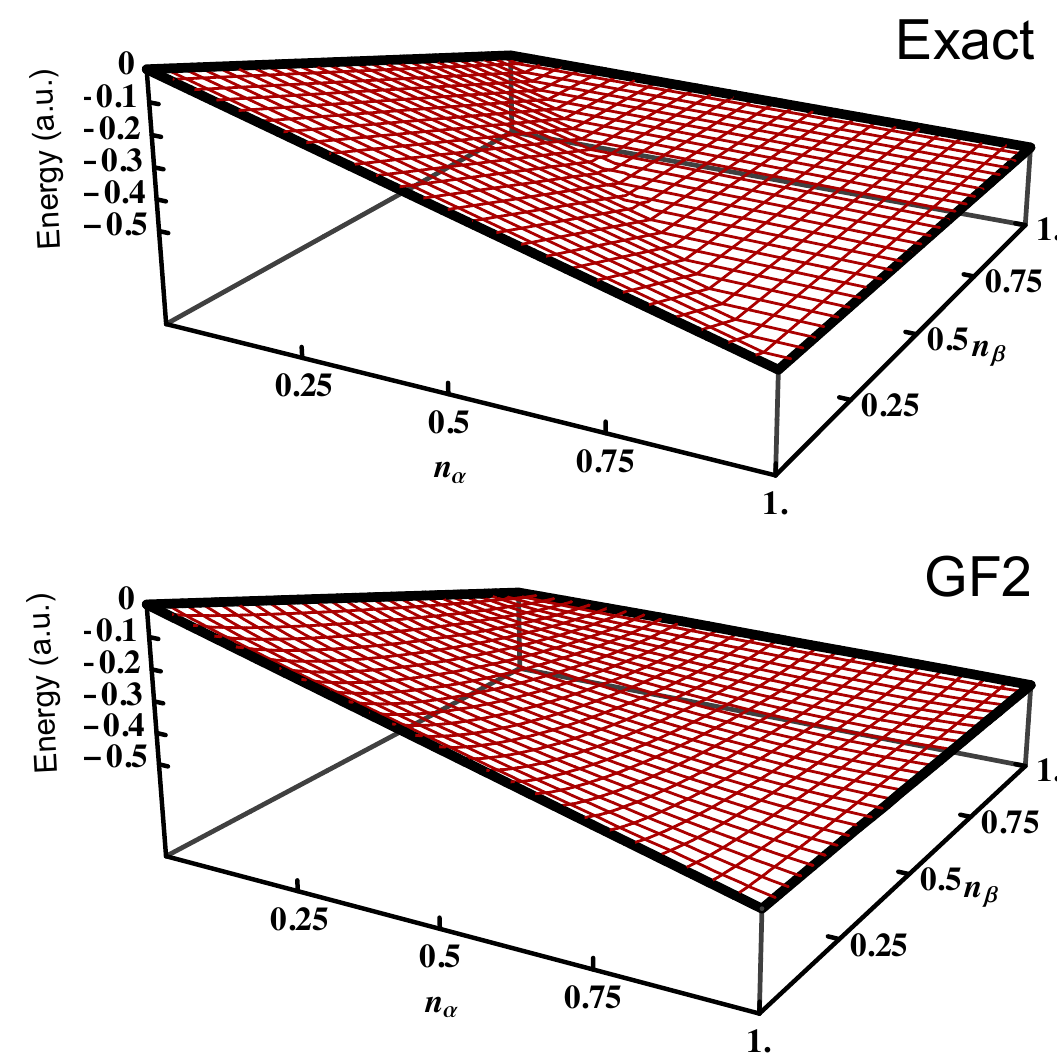}
\caption{The total energy for hydrogen with fractional spin up and down occupations $n_\alpha$, $n_\beta$ evaluated with  aug-cc-pVDZ. Top: the exact result. Bottom: GF2.}
\label{fig:hfrac}
\end{figure}

\begin{figure}
\includegraphics[width=8.5cm]{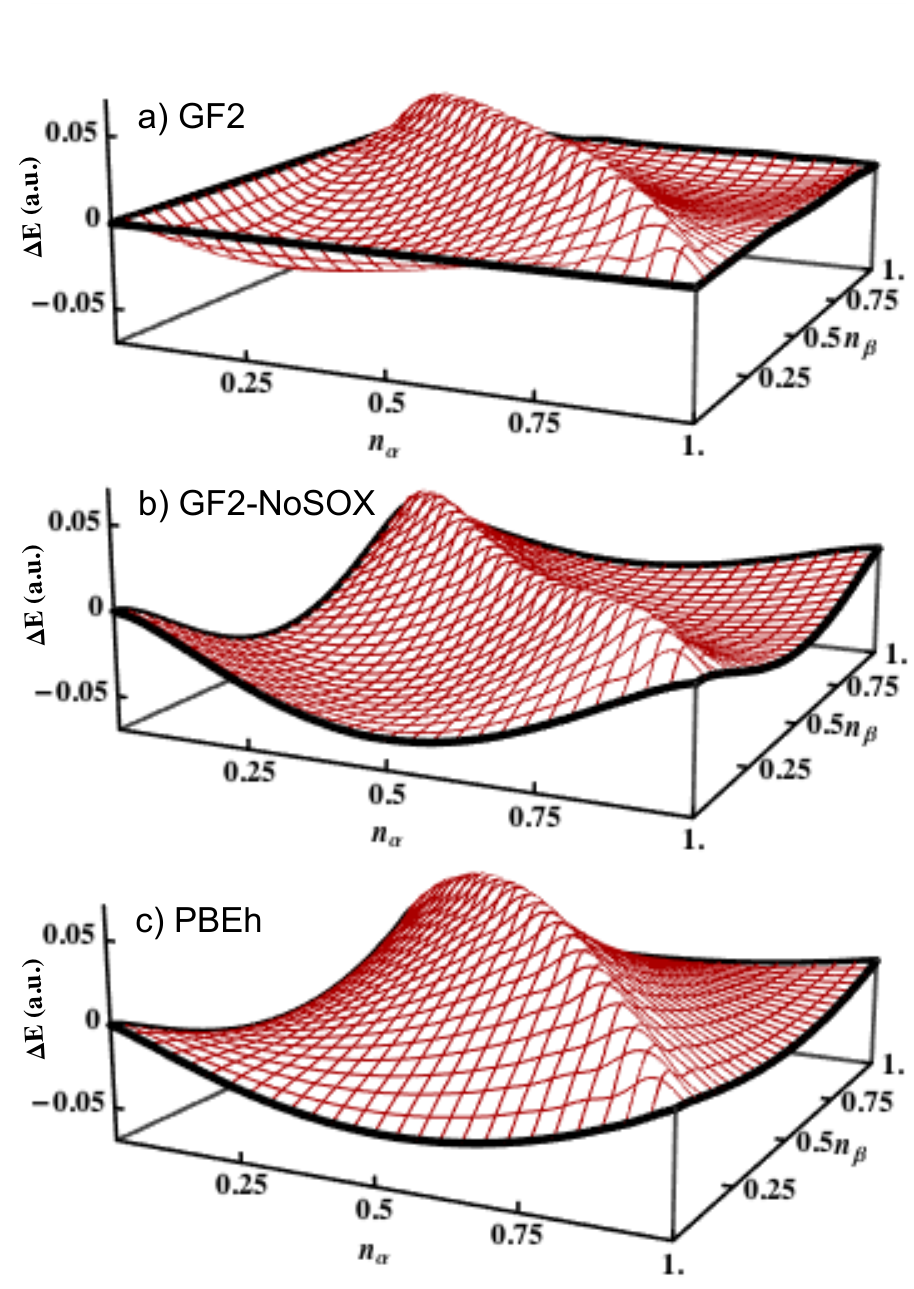}
\caption{The energy difference $\Delta E=E-E_{\textnormal{lin}}$ for the hydrogen atom with fractional spin up and down occupations $n_\alpha$, $n_\beta$, where $E$  is the energy evaluated with fractional electron number, and  $E_{\textnormal{lin}}$ is the flat-plane linear interpolation for a) GF2, b) GF2-NoSOX, and c) hybrid PBEh. All calculations are  with aug-cc-pVDZ}
\label{fig:diffarray}
\end{figure}

First we consider  a single hydrogen atom with fractional spin up and down electron occupations, $n_\alpha$, $n_\beta$, that are varied in the interval $0\le n_\alpha\le 1.0$ and $0\le n_\beta\le 1.0$. For this case with an exact method the energy should change linearly in the fractional electron number $n=n_\alpha +n_\beta $, while for  constant $n$ it should be invariant with respect to changes in the fractional spin $m=n_\alpha -n_\beta$. Furthermore, there should be a discontinuity in the slope $dE(n)/dn$ across the line $n_\alpha +n_\beta =1.0$. Hence the resulting energy surface should be two flat planes that intersect along a seam. In Figure~\ref{fig:hfrac} we show the exact result, compared against that obtained with spin-unrestricted GF2. If first we focus only on the edge of the plane where one occupation is held fixed at  integer values of 0 or 1.0, this corresponds to the fractional charge behavior of a method and thus SIE. It is clear that GF2 reproduces the exact linear behavior almost perfectly, similar to the result for MP2 in Ref.~\cite{Weitao_MP2_Frac_jctc2009}. Now if we focus on the more interesting region towards the interior, discrepancies between GF2 and the exact flat-plane behavior become apparent. To see this more clearly, in Figure~\ref{fig:diffarray} we plot the difference between GF2 and its interpolated flat-plane surface. Looking at the diagonal region connecting the coordinates $\{n_\alpha=1.0,~n_\beta=0.0\}$ and $\{n_\alpha=0.0,~n_\beta=1.0\}$ we find a hill of fractional spin error, where GF2 is not able to fully recover the static correlation energy. On either side of the hill ($n_\alpha +n_\beta <1.0$ and $n_\alpha +n_\beta >1.0$) we find shallow valleys where  GF2  moderately overestimates the correlation energy. The GF2 results here for fractional spin are in severe contrast to the MP2 result from Figure~4 of Ref.~\cite{Weitao_MP2_Frac_jctc2009}, which rapidly diverges to $-\infty$ correlation energy as one moves towards the center at $\{n_\alpha=0.5,~n_\beta=0.5\}$.

For purpose of comparison, we also show in Figure~\ref{fig:diffarray} the surface obtained with PBEh, as well as the GF2 result obtained when the second-order exchange (SOX) diagram is neglected in the self-energy (we call this GF2-NoSOX for short). Neglect of the SOX diagram introduces an unphysical one electron self-correlation error at second order. Because many electron SIE and static correlation are known to be connected\cite{PoloSIEcorrelation2002,SIE_and_StaticCorrelation_Henderson_2010,RPA_failure_Weitao_pra_2012,RPA_Renorm_BatesFurche_2013,BondBreak_MBPT_prl_2013}, GF2-NoSOX should provide an interesting contrast to standard GF2. First let us compare the energy landscapes for GF2 and  PBEh. Despite being a hybrid functional, PBEh still yields a significant fractional charge error along the outer rim of the surface, along with a pronounced hill of fractional spin error running through the center. The GF2 energy landscape in contrast is appreciably flatter, with less fractional spin and fractional charge error. This is notable, because it has been stressed that simultaneously reducing the fractional charge and spin errors in a single-particle method is very difficult\cite{RPA_failure_Weitao_pra_2012}. For example, RPA+X (RPA with  HF-type exchange) greatly improves over the fractional charge error of RPA, but at the price of gaining a considerably larger fractional spin error that is comparable to Hartree-Fock\cite{RPA_failure_Weitao_pra_2012}.
 Now let us examine the GF2-NoSOX result. As mentioned, neglect of the SOX diagram introduces a self-correlation error at second order, and as a result  GF2-NoSOX gives significant fractional charge errors comparable to PBEh. At the same time, the hill of fractional spin error is appreciably reduced relative to both PBEh and GF2.

\begin{figure}
\includegraphics[width=8.5cm]{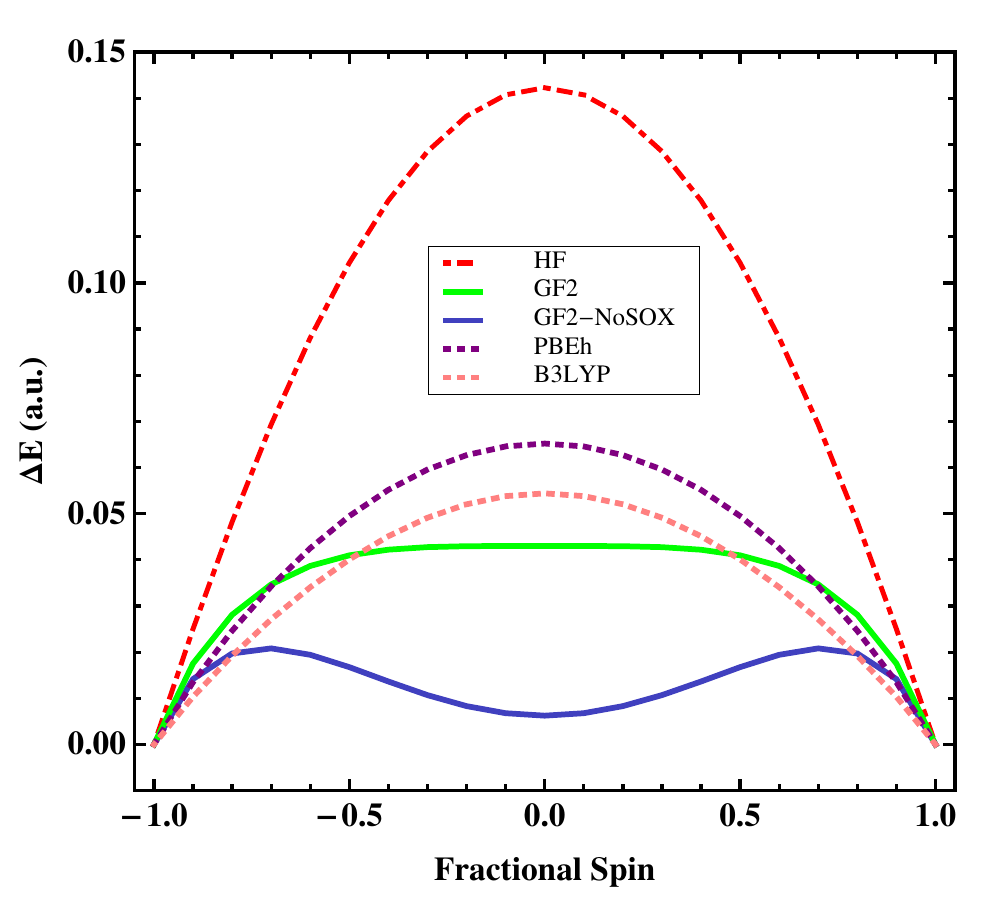}
\caption{The energy difference $\Delta E=E^{X}-E^{X}_{\textnormal{int}}$ for the fractional spin hydrogen atom, where $E^{X}$  is the energy from method $X$ evaluated with the fractional spin population $m=n_\alpha -n_\beta$, $n_\alpha + n_\beta =1$, and  $E^{X}_{\textnormal{int}}$ is the energy from  $X$ with integer  $m=\pm 1$. All calculations are  with aug-cc-pVDZ.}
\label{fig:fracspin}
\end{figure}

 To see the static correlation error more clearly, in Figure~\ref{fig:fracspin} we plot the energy with respect to fractional spin $m=n_\alpha -n_\beta$ for hydrogen with GF2, GF2-NoSOX,  PBEh,  B3LYP, and HF. This corresponds to the line running along the  energy surface from the coordinates $\{n_\alpha=1.0,~n_\beta=0.0\}$ to $\{n_\alpha=0.0,~n_\beta=1.0\}$. As stated previously an exact method should give a flat energy curve from $m=-1$ to $m=1$. For example, HF has a massive hill at $m=0$, which reflects the complete absence of static correlation energy in this method. Hybrid DFT yields much smaller fractional spin errors, with B3LYP being slightly lower than PBEh, likely because it includes less HF-type exchange than PBEh. In comparison the fractional spin error is relatively lower for GF2, as a result of the infinite order summation from the Dyson equation recovering some static correlation. GF2-NoSOX  has a much reduced fractional spin error relative to all four methods, with its energy at $m=0$ being not much different from that at $m=\pm 1$. It has been understood for some time that SIE can mimic static correlation\cite{PoloSIEcorrelation2002,SIE_and_StaticCorrelation_Henderson_2010,RPA_failure_Weitao_pra_2012,RPA_Renorm_BatesFurche_2013,BondBreak_MBPT_prl_2013}. Usually this is considered in the context of SIE resulting from incomplete cancellation of Coulomb and exchange terms at first order.  GF2-NoSOX's small fractional spin error in contrast is purely arising from incomplete cancellation of Coulomb and exchange terms at second order, resulting in an unphysical one-electron self-correlation that mimics static correlation energy, analogous to the situation that occurs with RPA+SOSEX\cite{SIE_and_StaticCorrelation_Henderson_2010} (second order screened exchange). We think this is an example of getting the right result for the wrong reason. Furthermore, GF2-NoSOX must obtain this slightly reduced fractional spin error  at the price of gaining a tremendous fractional charge error, which is not a desirable trade.
 In contrast,  from comparing Figures~\ref{fig:fracspin} and \ref{fig:diffarray} it is clear that GF2 has less static-correlation error than typical hybrid density functionals, and importantly achieves this while being essentially one and two electron self-interaction free. This means GF2 genuinely recovers some static correlation energy, rather than fortuitously exploiting spurious self-interaction or self-correlation. It is worth mentioning that a similar analysis has been performed for RPA and RPA+X\cite{RPA_failure_Weitao_pra_2012}. Comparing our GF2 result in Figure~\ref{fig:fracspin} to Figure~1 in Ref.\cite{RPA_failure_Weitao_pra_2012}, GF2 yields an appreciably smaller fractional spin error than RPA+X, but is still  larger than RPA. 


\begin{figure}
\includegraphics[width=8.5cm]{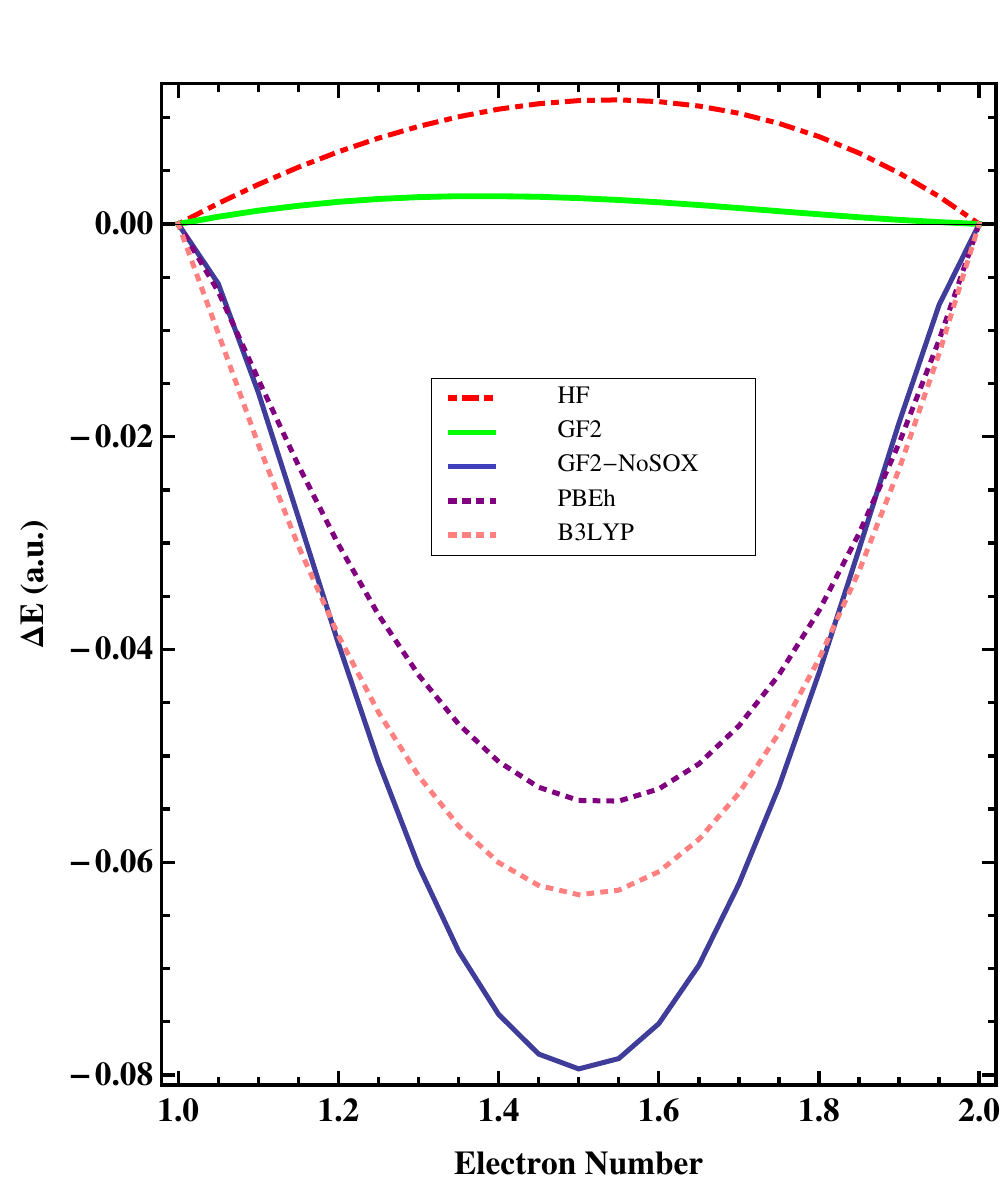}
\caption{The energy difference $\Delta E=E^{X}-E^{X}_{\textnormal{lin}}$ for the helium atom, where $E^{X}$  is the energy from method $X$ evaluated with fractional electron number, and  $E^{X}_{\textnormal{lin}}$ is the linear interpolation from method $X$. All calculations are  with cc-pVTZ. 
}
\label{fig:hefrac}
\end{figure}

Finally, to analyze the fractional charge error in closer detail  we consider the energy of helium with respect to fractional electron number. In Figure~\ref{fig:hefrac} we plot $E^{X}(n)-E^{X}_{\textnormal{lin}}$, where $E^{X}(n)$ is the energy from method $X$ for electron number $1.0\le n\le 2.0$, while $E^{X}_{\textnormal{lin}}$ is the linear-interpolation between integer points  with the same method. A method is said to be M-electron SIE free if $E^{X}(n)-E^{X}_{\textnormal{lin}}=0$ for $M-1\le n\le M$\cite{OneTwoManySIE2007}. We find that GF2 has a very small concave curvature. In contrast HF is moderately concave, while PBEh, B3LYP, and GF2-NoSOX are significantly convex. This clearly establishes that GF2 is one and almost perfectly two electron SIE free. Interestingly, comparing Figure~\ref{fig:hefrac} to Figure~4 in Ref.\cite{RPA_failure_Weitao_pra_2012}, it is reasonable to conclude that GF2 should have less SIE than both RPA and RPA+X. 

\section{Conclusions}

We have analyzed fractional electron errors in self-consistent Green's function theory by generalizing our previous GF2 implementation\cite{gf2_paper_2014} to open-shell systems. Overall we find that GF2 has a very small fractional charge error, and a moderate fractional spin error. In comparison to other well known methods, we find that GF2 has both less static correlation and self-interaction  error than hybrid density functionals such as B3LYP and PBEh, as well as RPA+X and HF. Because the GW approximation is diagrammatically identical to RPA, GF2 will very likely have significantly less self-interaction error than GW as well. Furthermore, it has been shown that CCSD has a roughly similar fractional charge error to  MP2\cite{WavefunctionFractional_Weitao_jcp_2013}. From this it stands to reason that GF2 and CCSD will have comparable fractional charge errors.

Essentially, by virtue of the Dyson summation GF2 greatly improves on the tremendous fractional spin error of MP2, but without deteriorating MP2's relatively excellent fractional charge behavior.  These results could suggest a way towards removing the fractional spin error from double hybrid density functionals\cite{DoubleHybrids_ijqc_2014,FracDoubleHybrid_Weitao_jpca_2014}.
As a further salient point, GF2  is fully self-consistent and thus the converged  density should reflect the relative lack of many-electron self-interaction error in a second order approximation. MP2 in contrast is by definition a perturbative scheme that does not revise the underlying mean-field reference, and thus inherits the Hartree-Fock density with its bias towards localization.  This suggests that GF2 could find good application for properties that sensitively depend on electron delocalization. For example, much of the interesting physics in transition metal complexes depends on the slight delocalization of unpaired $d$ electrons onto ligands, which is determined by the interplay of dynamic correlation  and  self-interaction error effects\cite{MetalLigDelocalizationMagOrbCabrero2002}.

Conceptually, GF2 is essentially a self-consistent single-particle theory where the energy is expressed as a functional of the single-particle Green's function $E[G]$, in obvious analogy to DFT with density functionals $E[\rho]$. In terms of the energy, it is fair to say GF2 is a ``Green's  function functional'' with desirable fundamental properties compared to standard  hybrid density functionals. 

\section{Acknowledgments }
J.J. P, A. A. K. and D.Z. acknowledge support by DOE ER16391.
J. J. P.  acknowledges Griffin C. Phillips for useful discussions.

\end{document}